\documentclass[12pt, oneside]{amsart}   	
\usepackage{geometry}              		
\usepackage{graphicx}				

\usepackage{amssymb}
\usepackage{bm}
\usepackage{booktabs}
\usepackage{hyperref}
\usepackage{amsaddr}
\usepackage{float}
\usepackage{mathtools}	

\title[Geometry of ZD]{The Geometry of Zero-Determinant Strategies}

\author{Xingru Chen$^{1, 2}$, Long Wang$^{3}$, and Feng Fu$^{2, 4}$}

\address{ 
$^1$School of Sciences, Beijing University of Posts and Telecommunications, Beijing 100876, China \\
$^2$Department of Mathematics, Dartmouth College, Hanover, NH 03755, USA \\
$^3$Center for Systems and Control, Peking University, Beijing 100871, China \\
$^4$Department of Biomedical Data Science, Geisel School of Medicine at Dartmouth, Lebanon, NH 03756, USA}

\email{xingrucz@gmail.com}
\email{longwang@pku.edu.cn}
\email{fufeng@gmail.com}

\date{}
						
\begin{document}
\maketitle

\section*{Abstract}
The advent of Zero-Determinant (ZD) strategies has reshaped the study of reciprocity and cooperation in the iterated Prisoner's Dilemma games. The ramification of ZD strategies has been demonstrated through their ability to unilaterally enforce a linear relationship between their own average payoff and that of their co-player. Common practice conveniently represents this relationship by a straight line in the parametric plot of pairwise payoffs. Yet little attention has been paid to studying the actual geometry of the strategy space of all admissible ZD strategies. Here, our work offers intuitive geometric relationships between different classes of ZD strategies as well as nontrivial geometric interpretations of their specific parameterizations. Adaptive dynamics of ZD strategies further reveals the unforeseen connection between general ZD strategies and the so-called equalizers that can set any co-player's payoff to a fixed value. We show that the class of equalizers forming a hyperplane is the critical equilibrium manifold, only part of which is stable. The same hyperplane is also a separatrix of the cooperation-enhancing region where the optimum response is to increase cooperation for each of the four payoff outcomes. Our results shed light on the simple but elegant geometry of ZD strategies that is previously overlooked. 

\section{Introduction}
The evolution of cooperation is one of the long-standing conundrums facing researchers from diverse fields~\cite{axelrod1981evolution}. Over the past decades, a number of seminal contributions have been made to improve our understanding of cooperation~\cite{nowak2006five}. Among others, cooperation can prevail under reciprocal altruism~\cite{trivers1971evolution}, which means ``you scratch my back and I will scratch yours.'' As a repeated two-player game, Iterated Prisoner's Dilemma games (IPD) have been the paradigm for studying direct reciprocity and cooperation~\cite{boyd1987no, nowak1990evolutionarily, nowak1992tit, nowak1993strategy, hauert1997effects, fudenberg2009folk, wu2018coevolutionary}. In particular, the famous Axelrold's tournament has ushered in an era of studying powerful IPD strategies using computer simulations in combination with analytical approaches~\cite{nowak1990stochastic, axelrod2012launching}. 

A plethora of memory-$n$ strategies including deterministic strategies and their stochastic counterparts have been thoroughly investigated~\cite{hauert1997effects,hilbe2017memory}. IPD strategies can be as sophisticated as possible, such as finite automata~\cite{rubinstein1986finite}, lookup table-based strategies~\cite{sandholm1996multiagent}, or strategies that are optimized by using techniques like neural networks or swarm particle intelligence~\cite{harper2017reinforcement}. On the other hand, winning IPD strategies can be surprisingly simple but powerful. The prominent Tit-for-Tat (TFT), for example, is such a ``fair-minded'' strategy that is found to be the backbone of direct reciprocity~\cite{nowak1992tit}. In the original version, TFT responds with full cooperation after co-player's cooperative move but always retaliates with full defection after co-player's defection. Variants of TFT, often called ``compliers''~\cite{hilbe2013evolution}, include the so-called generous TFT (GTFT)~\cite{nowak1993strategy} which forgives co-player's defection with a certain probability. These special strategies can be part of a larger class of reactive strategies that condition their responses on their co-player's behavioral choice (C vs D)~\cite{nowak1993strategy,baek2016comparing}. The notion of reactive strategies has enabled analytical insights into the evolution of reciprocity and cooperation in the IPD games~\cite{sigmund2010calculus}.

Memory-one IPD strategies are further specified by the probability to cooperate for the initial move and the probability to cooperate conditioned on each of the four possible payoff outcomes in a single round, denoted by $[q_1, q_2, q_3, q_4]$. The ordered labels $1$, $2$, $3$, and $4$, respectively, refer to the payoff outcome $R$ from the pair of strategy choices, $(C, C)$, $S$ from $(C, D)$, $T$ from $(D, C)$, and $P$ from $(D, D)$, as described from the focal row player's perspective via the payoff matrix
\begin{equation}
\begin{array}{cc}
   & C \quad D \\
  \begin{array}{c}
            C \\
            D
    \end{array}
   & \left( \begin{array}{cc}
              R & S \\
             T & P
             \end{array} \right).
\end{array}
\end{equation}
For Prisoner's Dilemma (PD), we have $T > R > P > S$. Conventional PD games also assume $2R > T+ S > 2P$, that is, mutual cooperation is better than unilateral cooperation better than mutual defection.

From the perspective of evolution, another strategy known as Win-Stay Lose-Shift (WSLS) stands out later~\cite{nowak1993strategy}. As for payoff control and manipulation, the discovery of equalizers is worthy of note~\cite{boerlijst1997equal}. The authors describe such a simple strategy, up to a properly chosen normalization factor $\phi$, $[ 1 - \phi(R - O), 1 - \phi(T - O), \phi(O - S), \phi(O - P)]$ that is able to set any co-player's average payoff $O$ to be the exact same amount between $P$ and $R$. Using an elegant approach of linear algebra~\cite{press2012iterated}, Press and Dyson further show that the so-called Zero-Determinant (ZD) strategies are able to unilaterally enforce a linear relationship between its average payoff $s_X$ and that of the co-player's $s_Y$, $s_X - O = \chi (s_Y - O)$, where $O$ is the baseline payoff, and $\chi$ is the extortion factor. Thus, the class of equalizers becomes a limiting subset of ZD strategies as $\chi \to \infty$ (that is, to unilaterally set co-player $Y$'s average payoff $s_Y$ to be $O$). ZD strategies can be categorized by their intended level of generosity $O$~\cite{stewart2013extortion}. The class with $O = P$ and $\chi  >1$ is often called extortionate ZD since these players can always ensure their advantage with an unfair surplus as $s_X - P = \chi (s_Y - P) \geq 0$ for conventional PD games. The class with $O = R$ and $\chi  >1$ is called generous ZD, ensuring that their own average payoff is never greater than the co-player's as $s_X - R = \chi (s_Y - R) \leq 0$. 

Efforts on various extensions to ZD strategies have been proven fruitful~\cite{mcavoy2016autocratic, ueda2021memory, knight2019recognising, govaert2020zero}, including but not limited to multi-person games~\cite{hilbe2014cooperation, pan2015zero, hilbe2015evolutionary, chen2022evolutionary},  noises or errors~\cite{hao2015extortion, mamiya2020zero}, and finitely repeated games~\cite{ichinose2018zero}. The evolution of ZD strategies has been studied in finite populations~\cite{hilbe2013evolution, hilbe2013adaptive} as well as in structured populations~\cite{szolnoki2014evolution}. The overall insight is that extortionate ZD strategies are powerful yet not evolutionarily stable in the sense that they will neutralize each other's advantage unless they adapt to be generous~\cite{adami2013evolutionary, chen2014robustness, hilbe2015partners, akin2016iterated}. Nevertheless, they can be the catalysts for the evolution of cooperation that pave the way for the emergence of cooperation (TFT, generous TFT, or more generally, generous ZD). Noteworthy, the dominance and optimality of ZD strategies depend on the underlying payoff structure, which is determined by the sign of $T + S - 2P$~\cite{chen2022outlearning}. It is found that a seemingly formidable ZD strategy can actually be outperformed, for example, by WSLS if $T + S < 2P$ and that when against fixed unbending strategies, the best response of ZD players is to offer a fair split by letting $\chi \to 1$~\cite{chen2022outlearning}.


Let us now turn to the superset of ZD strategies, which is a collection of memory-one strategies $[q_1, q_2, q_3, q_4]$ with three free parameters $(O, \chi, \phi)$:
\begin{equation}
\begin{split}
q_1 &= 1 - \phi(R - O)(\chi - 1), \\
q_2 &= 1 - \phi[(T - O)\chi + (O - S)], \\
q_3 &= \phi[(O - S)\chi + (T - O)], \\
q_4 &= \phi(O - P)(\chi - 1).
\end{split}
\label{ZDs}
\end{equation}
For $q_i$'s must be within $[0,1]$, the admissible ranges of $(O, \chi, \phi)$ are given by
\begin{equation*}
\left\{
\begin{array}{rcl}
P \le & O & \le R, \\
1 \le & \chi & < \infty ,  \\
0 \le & \phi & \le \min\{\frac{1}{(O - S)\chi + (T - O)}, \frac{1}{(T - O)\chi + (O - S)}\},
\end{array}
\right. 
\end{equation*}
or
\begin{equation*}
 \left\{
\begin{array}{rcl}
P \le & O & \le R, \\
-\infty \le & \chi & \le \min\{-\frac{O - S}{T - O}, -\frac{T - O}{O - S} \} \le -1,   \\
\max\{\frac{1}{(O - S)\chi + (T - O)}, \frac{1}{(T - O)\chi + (O - S)}\} \le & \phi & \le 0.
\end{array}
\right.
\end{equation*}

Geometrically, the family of memory-one ZD strategies formally expressed by a 4-dimensional tuple, the quadruple $[0,1]^4$, is located on a 3-dimensional hyperplane. Without loss of generality, it can be represented by:
\begin{equation}
q_4 = \frac{-(T + S - 2P)q_1 + (R - P)(q_2 + q_3) + T + S - R - P}{2R - T - S}.
\label{qlinear}
\end{equation} 
Prior work has almost exclusively focused on the resulting linear payoff relationship explicitly only with two parameters
$s_X - O = \chi(s_Y - O)$, assuming that player X uses a ZD strategy regardless of player Y's chosen strategy. While the ramification of ZD strategies can be conveniently visualized as a straight line in the parametric plot of $(s_X, s_Y)$, with the slope $1/\chi$ and the invariant point $(O, O)$, the actual geometry of ZD strategies is paid little attention. Having said so, the present work will shed new light on the elegance of ZD strategies in relation to their parameterizations $(O, \chi,  \phi)$ from the previously overlooked perspective of geometry.

\section{Results}

In this work, we will focus on the geometry of $(q_1, q_2, q_3)$ in the cube $[0, 1]^3$ for specific choices of $(O, \chi, \phi)$. Unless noted otherwise, we use the correspondence of Cartesian coordinates $(x,y,z)$ with respect to the ordered triplet $(q_1, q_2, q_3)$. There exists a one-to-one mapping, at least in the small neighborhood of a given ZD strategy, between $(q_1, q_2, q_3)$ and $(O, \chi, \phi)$ except for $\phi = 0$ or $\chi  = 1$, since the determinant of the Jacobian is 
\begin{equation}
\left|\frac{\partial(q_1, q_2, q_3)}{\partial(O, \chi, \phi)}\right| = (T - S)(2R - T - S)\phi^2(\chi - 1).
\end{equation}
For $\phi = 0$, the corresponding subset of ZD strategies regardless of the choice of $(O, \chi)$ degenerates into a point D $(1, 1, 0)$ (Fig.~1). Meanwhile, for $\chi = 1$, the corresponding subset of ZD strategies regardless of the choice of $O$ forms a line DA connecting $(1,0,1)$ and $(1, 1, 0)$ (Figs.~1a - 1c). This line is the common limit subset shared by all ZD strategies as $\chi \to 1$.

\begin{figure}[htbp]
   \centering
   \includegraphics[width=0.8\columnwidth]{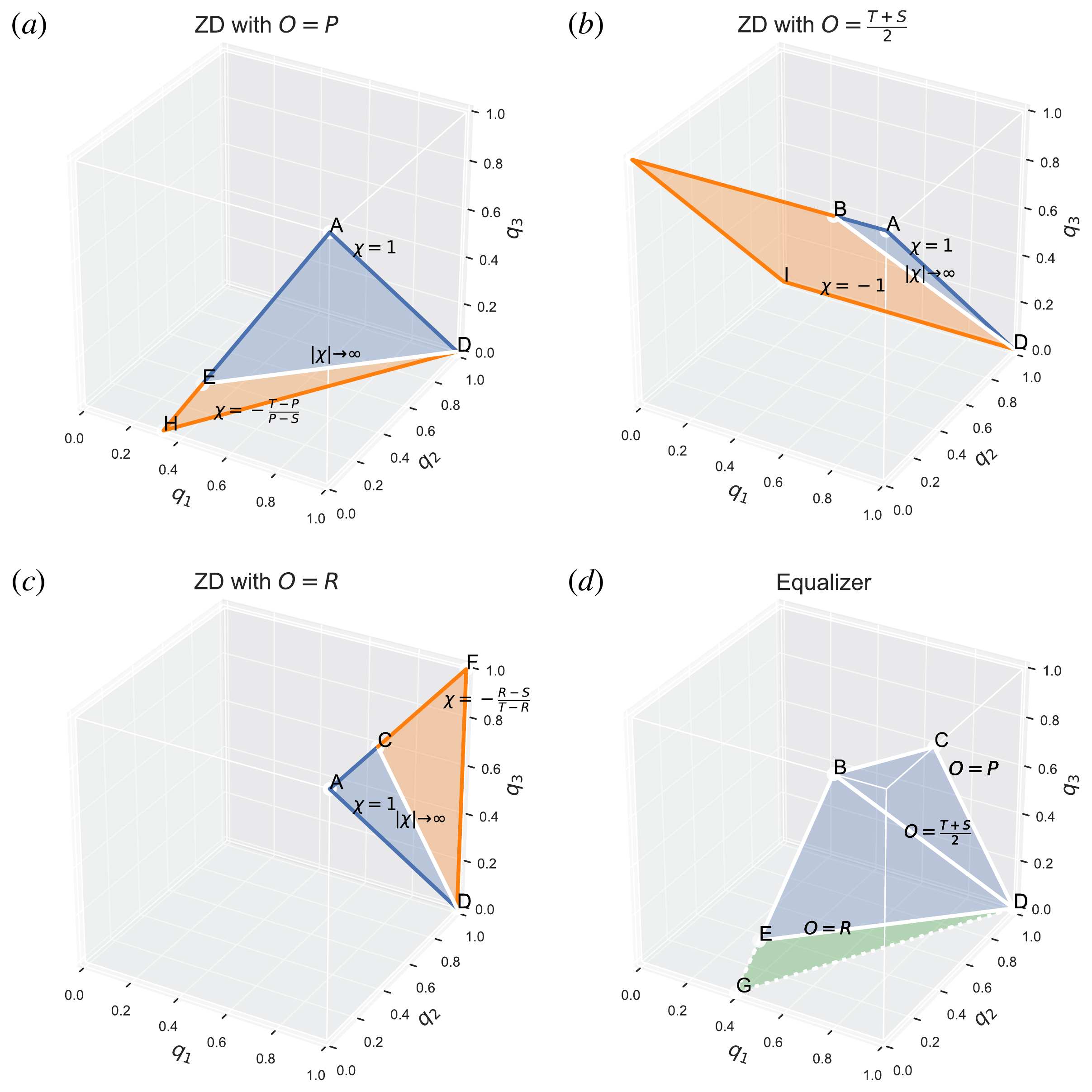} 
      \caption{Geometry of ZD strategies. Any ZD strategy $[q_1, q_2, q_3, q_4]$ can be visualized in the cube $[0, 1]^3$ for the ordered triplet $(q_1, q_2, q_3)$ using the fact that $q_4$ is linearly dependent on $q_1$, $q_2$, and $q_3$, as shown in Eq.~\eqref{qlinear}. (a), (b), and (c) show that ZD strategies with the same baseline payoff $O$ are on the same plane, respectively, for $O = P$, $(T+S)/2$, and $R$. (d) shows that all equalizer strategies are located on the same plane, indicated by BCDE, and the dashed triangle DEG indicates equalizer strategies that are inadmissible. The lines DE and DB and DC, also highlighted in (a) - (c), present equalizer strategies that are able to fix their co-player's average payoff to be $P$, $(T+S)/2$, and $R$, respectively. On the other hand, ZD strategies with different $O$ share the line DA as their common limit for $\chi \to 1$. When increasing $O$, the plane formed by ZD strategies with the same $O$ in effect rotates around the axis DA. All ZD strategies sharing the same $O$  and $\chi$ are on the same line; for a fixed $O$, the corresponding line approaches the limit of equalizer monotonically from both directions as $\chi$ tends to $+\infty$ from 1 or to $-\infty$ from $\min\{-(O - S)/(T - O), -(T - O)/(O - S)\}$, namely, from DA to DE and from DH to DE in (a), from DA to DB and from DI to DB in (b), and from DA to DC and from DF to DC in (c). For all figures in the paper, we consider the conventional payoff matrix $[R, S, T, P] = [3, 0, 5, 1]$.}
   \label{oplanes}
\end{figure}

For ZD strategies not in the line DA, their $(O, \chi, \phi)$ can be uniquely determined by $(q_1, q_2, q_3)$:
\begin{equation}
\begin{split}
O &= \frac{(T + S)q_1 - R(q_2 + q_3) + R - T - S}{2q_1 - q_2 - q_3 - 1}, \\
\chi &= \frac{(T - S)q_1 - (T - R)q_2 - (R - S)q_3 - (R - S)}{-(T - S)q_1 + (R - S)q_2 + (T - R)q_3 + T - R}, \\
\phi &= \frac{(T - S)q_1 - (R - S)q_2 - (T - R)q_3 - (T - R)}{(T - S)(2R - T - S)}.
\end{split}
\label{ZDpara}
\end{equation}

For ZD strategies with the same $O$, they all together form a 2-dimensional plane (Figs.~1a - 1c), given by
\begin{equation}
q_3 = \frac{(T + S - 2O)q_1 - (R - O)q_2 + R + O - T - S}{R - O}.
\label{Ofixed}
\end{equation}
The normal vector of this ``$O$-fixed plane'' is $\vec{n} = [T + S - 2O, -(R - O), -(R - O)]$. For instance, for baseline payoff $O = P$, the normal vector of this plane is $[T + S - 2P, -(R - P), -(R - P)]$ and for $O = R$, the normal vector is $[2R - T - S, 0, 0]$, or simply $[1, 0, 0]$.

Of particular interest, the ``equalizer'' subset of ZD strategies (obtained by letting $|\chi| \to +\infty$) forms another plane (Fig.~1d), which can be written as
\begin{equation}
q_3 = \frac{(T - S)q_1 - (R - S)q_2 - (T - R)}{T - R}.
\end{equation}
We note that the equation above is exactly equivalent to the denominator of the expression of $\chi$ as a function of $(q_1, q_2, q_3)$ given in Eq.~\eqref{ZDpara}. This plane of equalizer intersects with the plane of ZD strategies with fixed $O$, and the intersection line fixes any co-player's payoff to the same level $O$ (Fig.~1d).

For any given $O$, all of the corresponding ZD strategies contain the line DA connecting $(1, 1, 0)$ and $(1, 0, 1)$. In other words, the resulting plane formed by ZD strategies, while varying $O$ from $P$ to $R$, rotates with a common axis DA which is the line connecting $(1, 1, 0)$ and $(1, 0, 1)$. This unprecedented geometry of ZD strategies can also be indirectly validated by the fact that the normal vector of the plane $\vec{n} = [T + S - 2O, -(R - O), -(R - O)]$ formed by ZD strategies is always orthogonal to the vector DA $ = [0, -1, 1]$ for $P \le O \le R$.

To further explain the geometrical meaning of the parameters $(O, \chi, \phi)$, we now turn to specific examples of ZD strategies with $O = P$, $(T+S)/2$, $R$, respectively, as illustrated in Fig.~1. For ZD strategies on the same line that passes through any given triplet $(q_1, q_2, q_3)$ and $(1, 1, 0)$ (that is, point D), they all share the same $O$ and $\chi$. For fixed $O$, all the lines formed by such ZD strategies are on the same plane (Figs.~1a - 1c); when varying $\chi$, the resulting line rotates around $(1, 1, 0)$ towards a direction that depends on the sign of $\chi$ but has the exact same limiting position as $|\chi| \to +\infty$. For increasing positive $\chi$, the angle between this line and the vector DA $ = [0, -1, 1]$ monotonically increases with $\chi$ until reaching the limit of the ``equalizer" plane. For decreasing negative $\chi$, the line monotonically approaches the same ``equalizer'' line but from the other direction. In the following, we show this for positive $\chi$ values. The case for negative $\chi$ can be demonstrated analogously. 

The angle $\theta$ between the vector $[q_1 - 1, q_2 - 1, q_3]$ and $[0, -1, 1]$ is $\phi$-independent and given by 
\begin{align}
\small
\begin{split}
\cos \theta &= \frac{(q_1 - 1, q_2 - 1, q_3)\cdot(0, -1, 1)}{|(q_1 - 1, q_2 - 1, q_3)|\cdot|(0, -1, 1)|} \\
&= \frac{\sqrt{2}(T - S)(\chi + 1)}{2\{(R - O)^2(\chi - 1)^2 + [(T - O)\chi + O - S]^2 + [(O - S)\chi + T - O]^2\}^{1/2}}.
\end{split}
\end{align}
Its derivative with respect to $\chi$ is
\begin{equation}
\small
\frac{d\cos \theta}{d\chi} = -\frac{\sqrt{2}(T - S)(\chi - 1)[6O^2 - 4O(T + R + S) + 2R^2 + (T + S)^2]}{2\{(R - O)^2(\chi - 1)^2 + [(T - O)\chi + O - S]^2 + [(O - S)\chi + T - O]^2\}^{3/2}}.
\end{equation}
Because
\begin{equation}
\small
f(O) = 6O^2 - 4O(T + R + S) + 2R^2 + (T + S)^2 \geq f(\frac{T + R + S}{3}) = \frac{(2R - T - S)^2}{3} > 0,
\end{equation} 
we have $d\cos\theta/d\chi < 0$, suggesting that as $\chi$ increases, the line formed by ZD with the same $O$ and $\chi$ rotates around the point $(1, 1, 0)$ towards the equalizer limit line as $\chi \to +\infty$. For negative $\chi$, the corresponding line similarly rotates around the point $(1, 1, 0)$ towards the equalizer limit line as $\chi \to -\infty$.

Further, the equalizer plane separates all admissible ZD strategies into two regions: those above the plane with positive $\chi$ values whereas those below the plane with negative $\chi$ values (Fig.~2).

\begin{figure}[htbp]
   \centering
   \includegraphics[width=0.9\columnwidth]{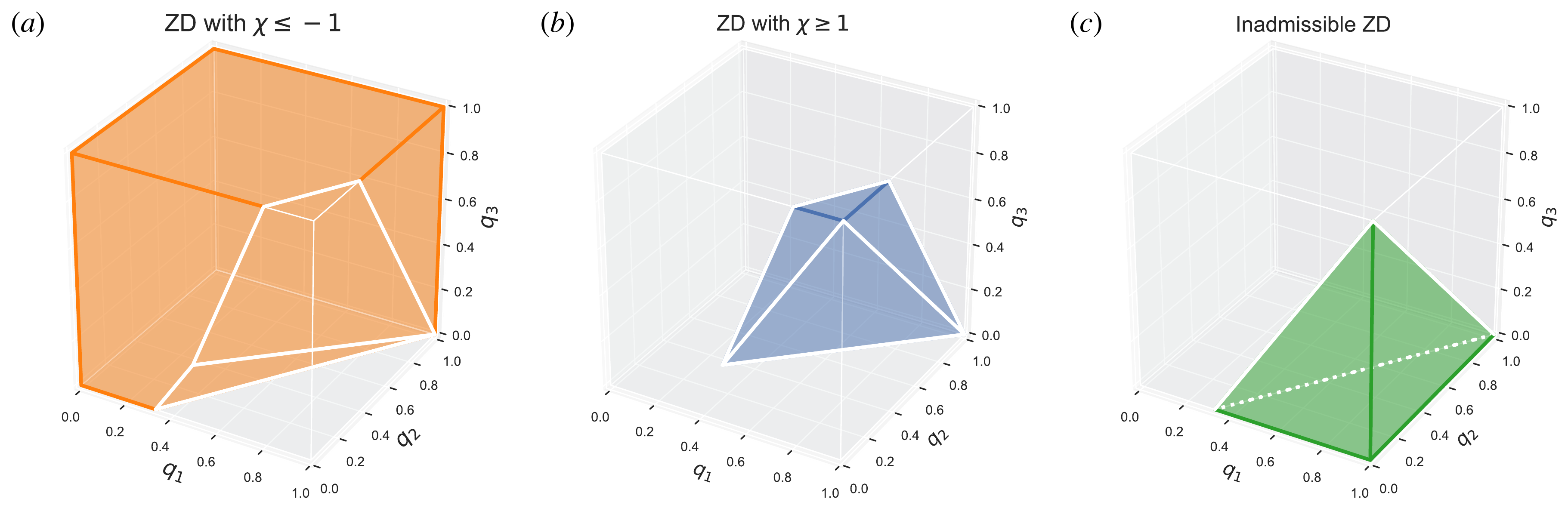} 
      \caption{Strategy space of admissible ZD strategies. The plane of equalizers separates the space of admissible ZD strategies into two regions: those with negative $\chi$ values in (a) and those with positive $\chi$ values in (b). The shaded region in (c) below the plane formed by ZD with $O = P$ (more specifically, triangle ADH in Fig.~1) and containing the corner $(1,0,0)$ of the cube is inadmissible as ZD strategies.}
   \label{gZD}
\end{figure}

Most strikingly, we find the nontrivial emergence of ``equalizers'' as the equilibrium manifold appearing in the adaptive dynamics of ZD strategies explicitly in the strategy space $(p_1, p_2, p_3)$ (Fig.~3a).

Assuming ZD strategies $\bm{p} = [p_1, p_2, p_3, p_4]$ vs $\bm{q} = [q_1, q_2, q_3, q_4]$, both satisfying Eq.~\eqref{qlinear}, Press and Dyson show that the average payoff $\pi(\bm{p},\bm{q})$ of player using $\bm{p}$ can be calculated as
\begin{equation}
\pi(\bm{p},\bm{q}) = \frac{\det \left |
\left.
\begin{array}{cccc}
 p_1q_1-1 & p_1 -1 & q_1-1 & R  \\
 p_2q_3& p_2 -1 & q_3 & S  \\
  p_3q_2 & p_3 & q_2-1 & T  \\
 p_4q_4 & p_4  & q_4 & P 
\end{array}
\right.
 \right|}{\det \left |
\left.
\begin{array}{cccc}
 p_1q_1-1 & p_1 -1 & q_1-1 & 1  \\
 p_2q_3& p_2 -1 & q_3 & 1  \\
  p_3q_2 & p_3 & q_2-1 & 1  \\
 p_4q_4 & p_4  & q_4 & 1
\end{array}
\right.
 \right|}
 .
\end{equation}
The adaptive dynamics of ZD strategies can be obtained by
\begin{equation}
\frac{dq_i}{dt} = \frac{\partial \pi(\bm{p},\bm{q})}{\partial p_i}\left|_{\bm{p} = \bm{q}} \right.
.
\end{equation}
After a bit of algebra, we get 
\begin{equation}
\begin{split}
\frac{dq_1}{dt} &= \frac{(q_2 + q_3 - 1)[(T - S)q_1 - (R - S)q_2 - (T - R)q_3 - (T - R)]}{(1 - q_2 + q_3)(1 - 2q_1 + q_2 + q_3)^2}, \\
\frac{d q_2}{dt} &  = \frac{1 - q_1}{q_2 + q_3 - 1}\cdot\frac{dq_1}{dt}, \\
\frac{d q_3}{dt} & =  \frac{d q_2}{dt}.
\end{split}
\end{equation}
We see that the selection gradient is zero at the equalizer plane, forming the equilibrium manifold of the corresponding adaptive dynamics (Fig.~3a). 

A simple calculation shows that 
\begin{equation}
(q_1 - 1)\frac{dq_1}{dt} + (q_2 - 1)\frac{dq_2}{dt} + q_3\frac{dq_3}{dt} = 0.
\end{equation}
Thus, the direction of the vector field $[dq_1/dt, dq_2/dt, dq_3/dt]$ is orthogonal to the vector $[q_1 - 1, q_2 - 1, q_3]$ that points from $(1, 1, 0)$ to $(q_1, q_2, q_3)$. As a matter of fact, the vector field is formed by the concentric spheres around the corner $(1, 1, 0)$.

\subsection{Stability of the equilibrium manifold}
The normal vector of the equalizer plane is $[T - S, -(R - S), -(T - R)]$. Its dot product with the selection gradient $[dq_1/dt, dq_2/dt, dq_3/dt]$ near the equilibrium manifold would be
\begin{equation}
\begin{split}
& [T - S, -(R - S), -(T - R)] \cdot [\frac{dq_1}{dt}, \frac{dq_2}{dt}, \frac{dq_3}{dt}] \\
 = & \frac{(T - S)(q_1 + q_2 + q_3 - 2)[(T - S)q_1 - (R - S)q_2 - (T - R)q_3 - (T - R)]}{(1 - q_2 + q_3)(1 - 2q_1 + q_2 + q_3)^2}.
\end{split}
\label{stability}
\end{equation}
The sign of this dot product can be used to determine the local stability of the equilibrium manifold. In particular, the plane $q_1+ q_2 + q_3 - 2 = 0$ intersects with the equalizer plane and divides it into stable and unstable parts (Fig.~3a). 

\begin{figure}[htbp]
   \centering
   \includegraphics[width=0.8\columnwidth]{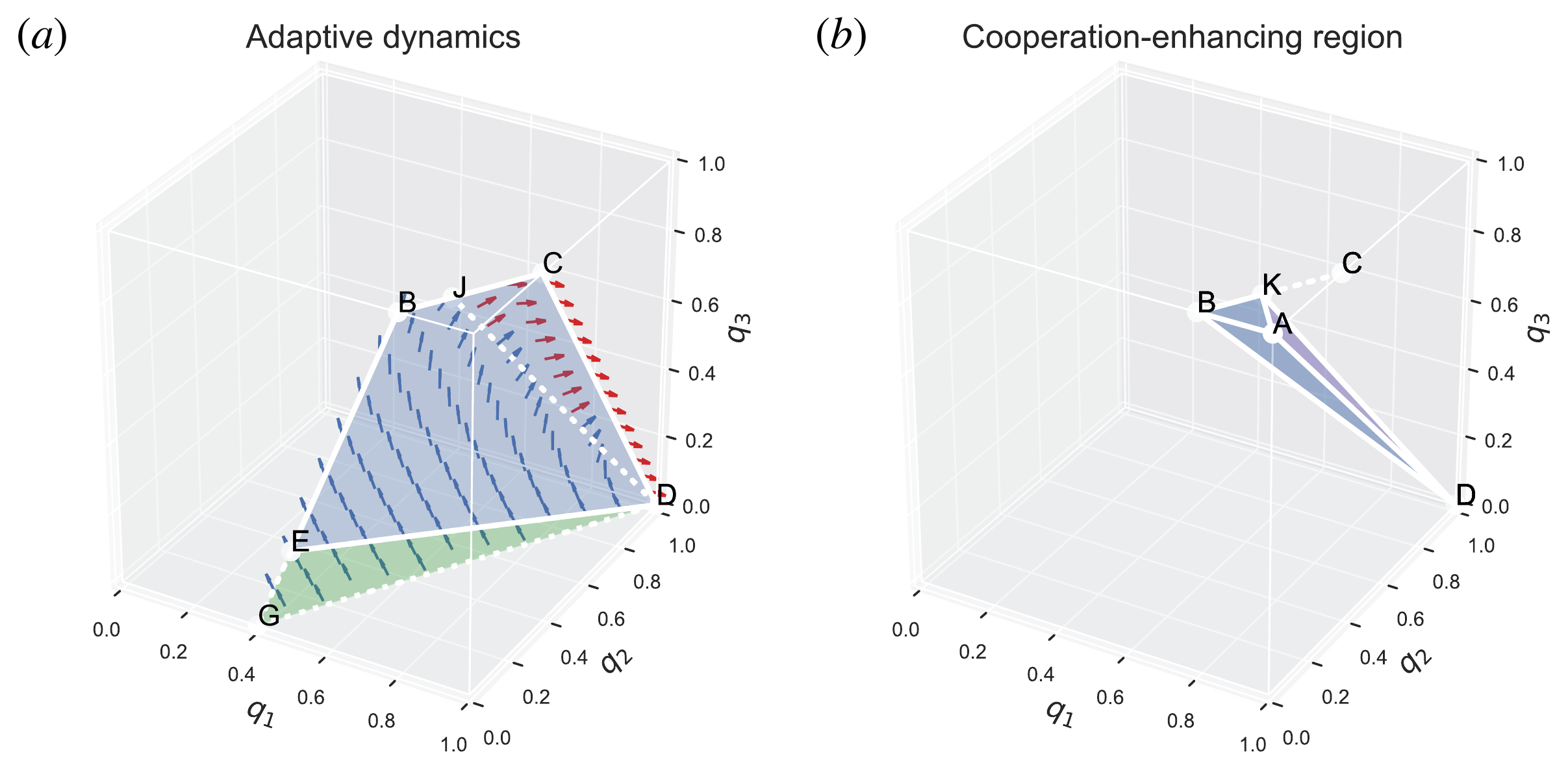} 
      \caption{Adaptive dynamics of ZD strategies. For clarity, (a) only shows the 3-dimensional vector field plot near the critical equilibrium manifold, which is the plane of equalizer strategies. Part of the equilibrium manifold, indicated by BEDJ, is stable and the remaining part, CDJ, is unstable. Different colors are used to indicate the sign of the dot product in Eq.~\eqref{stability}: blue for positive and red for negative. The line DJ is the intersection of the plane $q_1 + q_2 + q_3 - 2 = 0$ and the equalizer plane. The vector field $[dq_1/dt, dq_2/dt, dq_3/dt]$ is orthogonal to the vector $[q_1 - 1, q_2 - 1, q_3]$ that points from $(1, 1, 0)$ towards $(q_1, q_2, q_3)$. (b) shows the cooperation-enhancing region where $dq_i/dt$ is positive for  $i = 1, 2, 3, 4$ in adaptive dynamics. In this region, it is optimal for ZD strategies to increase their probabilities to cooperate invariably, after each of the four outcomes. }
   \label{adZD}
\end{figure}

\subsection{cooperation-enhancing region}
In the full strategy space, the cooperation-enhancing region can be found by requiring $\frac{dq_i}{dt} > 0$ for $i = 1, 2, 3, 4$. Using Eq.~\eqref{qlinear}, we get
\begin{equation}
\begin{split}
\frac{dq_4}{dt} & = \frac{1}{2R - S - T}\left[-(T + S - 2P)\frac{dq_1}{dt} + (R - P)(\frac{dq_2}{dt} + \frac{dq_3}{dt})\right] \\
& = \frac{2(R - P)(1 - q_1) - (T + S - 2P)(q_2 + q_3 - 1)}{(2R - T - S)(q_2 + q_3 - 1)}\cdot\frac{dq_1}{dt}.
\end{split}
\end{equation}
We thus obtain the critical plane $2(R - P)(1 - q_1) - (T + S - 2P)(q_2 + q_3 - 1) = 0$ so as to determine the sign of $dq_4/dt$. This plane is in fact the $O$-fixed plane (see Eq.~\eqref{Ofixed}) formed by ZD strategies with the specific $O$ value satisfying $(T+S)/2 < O < R$ and
\begin{equation}
\frac{2O - T - S}{R - O} = \frac{2(R - P)}{T + S - 2P}.
\end{equation}
Hence the cooperation-enhancing region (as shown in Fig.~3b) is given by
\begin{equation}
\begin{split}
&q_2 + q_3 - 1 > 0,\\
&(T - S)q_1 - (R - S)q_2 - (T - R)q_3 - (T - R) > 0, \\
&2(R - P)(1 - q_1) - (T + S - 2P)(q_2 + q_3 - 1) > 0.
\end{split}
\end{equation} 

\subsection{Equal gains from switching: \texorpdfstring{$T + S = R + P$}{}}

A necessary condition for reactive strategies to be ZD strategies is that the four payoff values need to satisfy the so-called equal gains from switching: $T + S = R + P$. Under this condition, we may further get $q_1 = q_3$ and $q_2 = q_4$.

By letting $q_1 = q_3$ and $q_2 = q_4$, we obtain $\phi =1/[(R - S)\chi + T - R]$ and $\phi= 1/[(T - P)\chi + P - S]$, respectively. Moreover, equating these two $\phi$ values yields
\begin{equation}
\frac{(T + S - R - P)(\chi - 1)}{[(R - S)\chi + T - R][(T - P)\chi + P - S]} = 0.
\end{equation}
Hence, the payoff condition $T+S = R + P$ is required as desired. To obtain the reactive strategy $[q_1, q_2, q_1, q_2]$ properly from the superset of ZD strategies, we also need to employ the following combinations of $(O, \chi)$:
\begin{equation}
\begin{split}
O &= \frac{(T + S)(1 - q_1) + R(q_1 + q_2 - 1)}{1 - q_1 + q_2}, \\
\chi &= \frac{R - S - (T - R)(q_1 - q_2)}{(R - S)(q_1 - q_2) - (T - R)}.
\end{split}
\end{equation}

Under this payoff structure condition, another interesting finding is that unconditional strategies ($q_1 = q_2 = q_3 = q_4$) are ZD strategies with negative $\chi$. Using Eq.~\eqref{qlinear} and forcing $q_1 = q_2 = q_3$, we get
$q_4 = q_1 + (T + S - R - P)/(2R - T - S)$, which leads to the same ``equal gains from switching'', $T + S = R + P$, in order for $q_4 = q_1$. The corresponding choice of $O, \chi, \phi$ are now 
\begin{equation}
\begin{split}
O &= (2R - T - S)q_1 + T + S - R, \\
\chi &= -\frac{R - S}{T - R}, \\
\phi &= -\frac{T - R}{(T - S)(2R - T - S)}.
\end{split}
\end{equation}

\subsection{Fun facts about ZD strategies}
The term $2q_1  - q_2 - q_3 -1$ is the denominator of the expression of $O$ as a function of $(q_1, q_2, q_3)$ in Eq.~\eqref{ZDpara}. For all admissible ZD strategies, we actually have
\begin{equation}
2 q_1  - q_2 - q_3 -1 < 0.
\end{equation}
Algebraically, using Eq.~\eqref{ZDs}, we get $2q_1 - q_2 - q_3 - 1 = -(2R - T - S)\phi(\chi - 1) < 0$ for any admissible parameter choices. The same inequality can also be shown by utilizing the exquisite geometry of ZD strategies. As $O \to -\infty$, the $O$-fixed plane approaches the limit $q_3 = 2q_1 - q_2 -1$, which is further below the plane for $O = P$.  Likewise, as $O \to +\infty$, the $O$-fixed plane rotates further beyond the plane corresponding to $O = R$ (whose equation is $q_1 = 1$) and approaches the same limit from the other direction. Since a ZD strategy only admits $P\le O \le R$, it follows that $q_3 > 2q_1 - q_2 - 1$ for all ZD strategies. Hence we obtain the inequality above as desired. 

\section{Discussions and Conclusion}

In this paper, we focus on the payoff structure of conventional PD games satisfying $T+S > 2P$. It is not impossible that in some PD games $T + S  < 2P$ holds, that is, the polygon connecting $(R,R)$, $(S,T)$, $(P,P)$, and $(T,S)$ is non-convex in the pairwise payoff plot~\cite{chen2022outlearning}. Under this circumstance of more adversarial nature, the geometry of Zero-Determinant (ZD) strategies can be studied similarly. It is also worth noting that, for games satisfying ``equal gains from switching'', $T + S = R + P$, reactive strategies (including unconditional strategies) become special cases of ZD strategies. 

Scrutinizing the geometry of ZD strategies provides useful intuition about and insights into how ZD strategies relate to one another. The parameterization combination $(O, \chi, \phi)$ of ZD strategies determine the admissible range of $\bm{q} = [q_1, q_2, q_3, q_4]$ that can be realized. Each subset of ZD strategies with a fixed $O$ value contains a common extreme boundary AD (Figs.~1a - 1c) that is a linear interpolation of TFT ($\bm{q} = [1, 0, 1, 0]$) and ``AllC or AllD'' ($\bm{q} = [1,1, 0, 0]$). The line formed by ZD strategies sharing the same $(O, \chi)$ approaches the limit of equalizer strategies, located on the plane BCDE (Fig.~1d), as $\chi \to +\infty$ or $\chi \to -\infty$.  This equalizer plane in fact separates the strategy space $(q_1, q_2, q_3)$ of admissible ZD strategies into two regions with positive versus negative $\chi$ values (Figs.~2a - 2b). We also realize that the region of the cube under the plane expanded by ZD strategies with $O = P$ and containing the corner $(1, 0, 0)$ is inadmissible for any ZD strategies (Fig.~2c).

From the perspective of evolution, it is straightforward to use the adaptive dynamics of ZD strategies to figure out the potential optimal ZD strategies against each other. We find that the common line AD of all ZD strategies with $\chi  \to 1$ emerges as a particular equilibrium. More generally, the plane of equalizers is the non-trivial equilibrium manifold, part of which is stable whereas the remaining part is not (Fig.~3a). The same plane acts as a separatrix of the cooperation-enhancing region that encourages increases in cooperation for each $q_i$ where $i = 1, 2, 3, 4$ (Fig.~3b).

In sum, this paper reveals the non-trivial and elegant geometry of ZD strategies that is previously overlooked. Our present work offers a geometrical interpretation of the ZD parameters $(O, \chi, \phi)$ and particularly the unprecedented geometrical relationships of ZD strategies in the entire strategy space of memory-one IPD strategies with $\bm{q} = [q_1, q_2, q_3, q_4]$. Most interestingly, we find that the subset of equalizer strategies forms the critical plane in the strategy space that emerges as the critical equilibrium manifold in the adaptive dynamics of ZD strategies and also as a separatrix of the cooperation-enhancing region. These results highlight the previously unforeseen connection between equalizers and general ZD strategies.

\section{Appendix}

The symbolic coordinates of the points in all the figures are given below. Here, $d_k = (T - P)^2 + (R - P)^2 + (R - S)^2 - 2(P - S)^2$.

\begin{table}[H]
\begin{tabular}{l l l l l}
A: $(1, 0, 1)$ & & B: $(\frac{2(T - R)}{T - S}, 0, 1)$ & & C: $(1, \frac{2R - T - S}{R - S}, 1)$ \\
& & & & \\
D: $(1, 1, 0)$ & & E: $(\frac{T - R}{T - P}, 0, \frac{P - S}{T - P})$ & & F: $(1, 1, 1)$ \\
& & & & \\
G: $(\frac{T - R}{T - S}, 0, 0)$ & & H: $(\frac{T + S - R - P}{T + S - 2P}, 0, 0)$ & & I: $(0, 1, 0)$ \\
& & & & \\
J: $(\frac{2T - R - S}{T + R - 2S}, \frac{2R - T - S}{T + R - 2S}, 1)$ & & 
\multicolumn{3}{l}{K: $(1 - \frac{(T + S - 2P)(2R - T - S)}{d_k}, \frac{2(R - P)(2R - T - S)}{d_k}, 1)$} \\ 
& & & & \\
\end{tabular}
\end{table}

In particular, if we consider the conventional Prisoner's Dilemma game with payoff matrix $[R, S, T, P] = [3, 0, 5, 1]$, we further get the following specific numbers for all the figures.

\begin{table}[H]
\begin{tabular}{l l l l l l l}
B: $(\frac{4}{5}, 0, 1)$ & & C: $(1, \frac{1}{3}, 1)$ & & E: $(\frac{1}{2}, 0, \frac{1}{4})$ & & G: $(\frac{2}{5}, 0, 0)$ \\
& & & & & & \\
H: $(\frac{1}{3}, 0, 0)$ & & J: $(\frac{7}{8}, \frac{1}{8}, 1)$ & & K: $(\frac{8}{9}, \frac{4}{27}, 1)$ & & \\ 
& & & & & & \\
\end{tabular}
\end{table}

\subsection{Data avaiability} All data pertaining to the present work has been included in the paper.

\subsection{Code avaiability} The Python Jupyter notebook that can be used to reproduce results in the paper is available at Github: https://github.com/fufeng/ZDGeometry.

\end{document}